\documentclass[fleqn,10pt]{wlscirep}
\usepackage[utf8]{inputenc}
\usepackage[T1]{fontenc}
\usepackage{xcolor}

\title{Replica Symmetry Breaking in the Quasi Mode Locked Regime of Titanium Sapphire Lasers}

\author[1,2,3]{Nicolas P. Alves}
\author[1]{Ricardo V. M. de Almeida Filho}
\author[1]{Isaac C. Nunes}
\author[1,4]{Gleison S. Bezerra}
\author[1]{Cid B. de Araújo}
\author[1]{Marcio H. G. de Miranda}
\author[1, *]{André C. A. Siqueira}

\affil[1]{Departamento de F\'{\i}sica, Universidade Federal de Pernambuco, Recife-PE, 50670-901, Brazil}
\affil[2]{European Laboratory for Nonlinear Spectroscopy (LENS), 50019 Sesto Fiorentino, Italy}
\affil[3]{Department of Physics and Astrophysics, University of Florence, 50019 Sesto Fiorentino, Italy}

\affil[4]{Programa de Pós-Graduação em Ciência de Materiais, Universidade Federal de Pernambuco, 50670-901 Recife-PE, Brazil}

\affil[*]{ andre.chaves@ufpe.br}

\begin{abstract}
Recent observations of replica symmetry breaking (RSB) in ytterbium-based mode-locked fiber lasers have revealed glassy behavior in the quasi-mode-locked (QML) regime. 
Here, we investigate whether this behavior is restricted to a specific laser implementation or represents a more general feature of mode-locked systems.
By analyzing statistical correlations of spectral intensity fluctuations through the Parisi overlap and Pearson correlation parameters, we study two Titanium:sapphire laser systems: a homemade cavity and a commercial laser. 
In both cases, clear signatures of RSB are observed in the QML regime, while replica-symmetric behavior is recovered in the amplified spontaneous emission (ASE) and standard mode-locked (SML) regimes. 
These results demonstrate the occurrence of RSB in Titanium:sapphire lasers and support the interpretation of the QML regime as a photonic glassy phase. Together with previous observations in ytterbium-based fiber lasers, these findings suggest that the connection between QML dynamics and RSB may be a general feature of mode-locked laser systems.

\end{abstract}

\begin{document}
    
\flushbottom
\maketitle
%
%
\thispagestyle{empty}

\section*{Introduction}

In laser systems supporting many interacting modes, nonlinear coupling, competition for gain, and intrinsic fluctuations may give rise to complex statistical regimes.
In these regimes, reproducibility at the individual level is lost, while well-defined statistical properties emerge. 
Such behavior naturally invites a description in terms of statistical physics, where concepts originally developed for disordered systems have proven useful for identifying and classifying collective photonic states \cite{ref01, ref02, ref03, ref04, ref05, ref06, ref07, ref08, ref09, ref10}. 
Within this framework, replica symmetry breaking (RSB) provides a powerful tool to characterize such regimes. 
Originally introduced in the theory of magnetic spin glasses, RSB describes situations in which systems prepared under identical conditions evolve toward inequivalent configurations due to competing/frustrated interactions. 
A key parameter in this analysis is Parisi's overlap parameter $q$ and its probability distribution $P(q)$, which quantifies correlations among independent replicas of a system \cite{ref08, ref09, ref10, ref11, ref12, ref13}. 
Here, $q$ is defined as the degree to which two independent replicas (configurations) of the system prepared under the same thermodynamic conditions are correlated.

The experimental validation of RSB was first achieved in random lasers (RLs), establishing the magnetic-to-photonic analogy \cite{ref14, ref15, ref16, ref17, ref18, ref19, ref20, ref21}. 
In RLs, optical feedback arises from multiple scattering due to refractive-index inhomogeneities in the gain medium, leading to intensity fluctuations that map onto the \textit{paramagnetic-to-spin-glass phase transition} of magnetic systems. 
In the photonic case, the concept of replicas corresponds to the spectra emitted at identical initial conditions, and the pump power plays the role of the inverse temperature.
The overlap parameter $q_{\alpha\beta}$ between two spectra $\alpha$ and $\beta$ then measures the normalized cross-correlation of their intensity fluctuations across spectral modes \cite{ref14, ref15}.

Below the RL threshold, uncorrelated fluctuations in emitted fluorescence produce a distribution $P(q)$ centered at $q=0$, consistent with a replica-symmetric phase characteristic of the paramagnetic state. 
Above the threshold, increasing mode competition and frustration drive the system into a replica symmetry broken phase, where $P(q)$ develops a bimodal structure with maxima near $q\approx\pm 1$, characteristic of a photonic spin-glass phase. 
Moreover, at sufficiently high excitation, the distribution may progressively return to a centered $q=0$ distribution as the system approaches a more ordered regime \cite{ref14, ref15}. 
In this case, gain saturation and mode competition drive the selection of a reduced set of dominant modes, effectively suppressing fluctuations among spectral replicas. 
As a result, the system exhibits more stable and reproducible spectral configurations, consistent with a return to replica-symmetric behavior.

Beyond RLs, signatures of RSB have been reported in a wide range of photonic platforms \cite{ref22_0}, including disordered nonlinear media \cite{ref22}, filamentation phenomena \cite{ref23, ref24, ref25}, higher-order soliton dynamics \cite{ref26}, Q-switched Nd:YAG lasers \cite{ref27}, and, more recently, mode-locked fiber lasers \cite{ref28, ref29}. 
Despite the diversity of underlying physical mechanisms, these systems exhibit remarkably similar statistical features, suggesting that RSB provides a unifying framework for describing complex photonic dynamics.
Additionally, the Pearson correlation coefficient has been employed as a complementary quantity, showing interactions among modes within a single spectrum. 
It provides valuable information on the development of mode competition in the gain medium \cite{ref28, ref29, ref30}.

A particularly interesting scenario emerges in mode-locked fiber lasers operating in the quasi-mode-locked (QML) regime. 
The QML regime sits between the low-excitation regimes, typically continuous wave (CW), and the standard mode-locked (SML) operation at higher pump levels. 
It is characterized by strong fluctuations in intensity and partial mode organization, without any global phase locking. 
Furthermore, its spectral bandwidth is narrower than that observed in both the CW and SML regimes. 
Recent experimental investigations of an ytterbium-doped fiber laser have revealed clear signatures of RSB in this intermediate regime \cite{ref28}, pointing to the existence of a glassy phase associated with frustrated nonlinear interactions among longitudinal modes.
The identification of such a glassy phase provides a means of understanding the complex interactions that govern the dynamics of mode-locked lasers before stable pulse formation.
However, it remains unclear whether this glassy behavior is confined to this specific system or represents a more general feature of multimode laser systems.

In this work, we address this topic by analyzing the statistical properties of the QML regime in Titanium:sapphire (Ti:sapphire) lasers. 
These systems provide a markedly different physical platform from fiber-based lasers, featuring a broadband gain medium, Kerr-dominated nonlinearity, and short cavity lengths \cite{ref31, ref32, ref33, ref34, ref35, ref36}. 
By computing $P(q)$ and the Pearson correlation parameter from ensembles of sequentially acquired spectral realizations recorded under stationary operating conditions for both a homemade cavity and a commercial laser (Coherent Mira Optima 900-F), we identify clear signatures of RSB in the QML regime of Ti:sapphire lasers, confirming that the glassy QML regime is not restricted to fiber-based systems.

\section*{Results}

\subsection*{Regimes of Ti:sapphire lasers}

Different classes of laser systems exhibit distinct physical mechanisms in the low-excitation regime: random lasers emit fluorescence below the lasing threshold, and ytterbium-based mode-locked fiber lasers typically exhibit CW emission. 
In Ti:sapphire lasers, low-excitation regions are characterized by the presence of amplified spontaneous emission (ASE), with a very large optical bandwidth. 
Despite these differences, the underlying statistical properties in these low-excitation regimes are remarkably similar across systems. 
The lack of long-range correlations leads to Gaussian distributions in the intensity fluctuations. 
Within the statistical framework of RSB, this regime corresponds to a replica-symmetric phase, analogous to a paramagnetic state in magnetic systems, in which spins are randomly oriented due to negligible spin-spin interactions. 

\begin{figure}[ht!]
\centering
\includegraphics[width=88 mm]{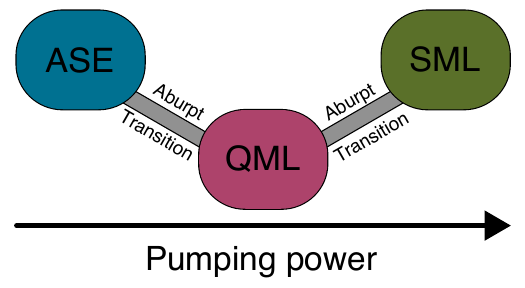}
\caption{Schematic optical phase diagram as a function of pump power for a Ti:sapphire laser, illustrating the transition from low-excitation ASE operation to the intermediate QML regime and, at higher pump powers, to the SML regime. The corresponding magnetic analogs are the paramagnetic, RSB glassy, and replica-symmetric ferromagnetic phases.}
\label{fig:fig_1}
\end{figure}

In mode-locked laser systems, increasing the pump power drives the system away from the low-excitation regime toward the QML state. 
Here, optical nonlinearity starts to play an increasingly relevant role, and additional longitudinal modes become active in the cavity. 
These modes compete for gain and exhibit strong intensity fluctuations, giving rise to glassy-like dynamics and the emergence of RSB. 
At higher pump powers, under appropriate cavity alignment and operating conditions, the system reaches the SML regime, characterized by the formation of stable ultrashort pulses and global phase locking among temporal modes. 
The dynamics in this regime are replica-symmetric and associated with a ferromagnetic-like phase, in which spins are aligned in the same direction without frustration \cite{ref27, ref28, ref29}.

\begin{figure}[ht!]
\centering
\includegraphics[width=\textwidth]{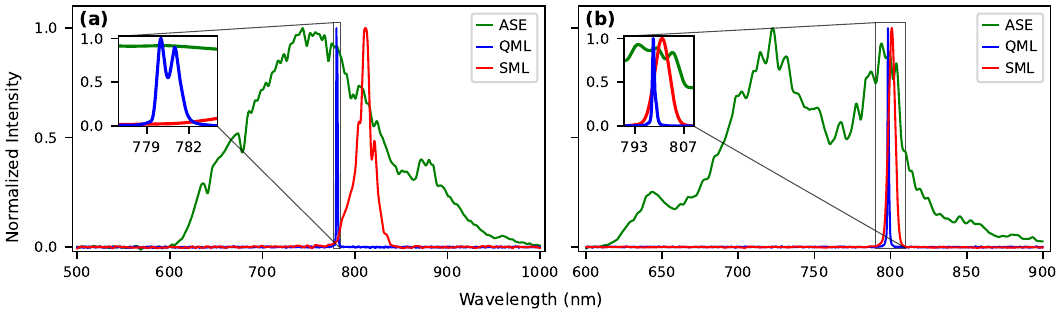}
\caption{Optical spectra for the (a) homemade and (b) commercial cavities. An inset in the top left shows the underlying structure of the spectrum in the QML region, revealing a narrow profile with the emergence of single or double peaks. Spectra are normalized for better visualization. In the homemade cavity, pump powers were $0.5$, $2.5$, and $5.0$ W for the ASE, QML, and SML regimes, respectively; for the commercial cavity, they were $2.0$, $3.7$, and $5.0$ W.}
\label{fig:fig_2}
\end{figure}

These behaviors are summarized in the general optical phase diagram shown in Fig. \ref{fig:fig_1}, which displays three distinct regimes (ASE, QML, and SML) as pump power increases. 
The transitions between regimes are marked by changes in the spectral profile, reflecting a reorganization of the active modes at each point. 
The corresponding magnetic analogies follow the theoretical framework developed for photonic systems \cite{ref13}. 
Here, we focus on the transition into the QML regime to examine whether RSB is a robust and universal intrinsic feature of this phase in Ti:sapphire laser systems.
Since the statistical analysis is based on the spectral intensity fluctuations, we first need to characterize the optical spectra across the distinct dynamical regimes. 
Fig. \ref{fig:fig_2} shows representative optical spectra obtained from the homemade Ti:sapphire laser (Fig. \ref{fig:fig_2}(a)) and the commercial Mira (Fig. \ref{fig:fig_2}(b)). 
The detailed setup of both systems is described in the Experimental Details section. 
The overlaid spectra correspond to the three dynamical regimes identified in the phase diagram of Fig. \ref{fig:fig_1}: the ASE regime in green, the intermediate QML regime in blue, and the SML regime in red. For the homemade cavity, the corresponding pump powers were $0.5$, $2.5$, and $5.0$ W, respectively; for the commercial cavity, they were $2.0$, $3.7$, and $5.0$ W.

In the ASE regime, the spectrum consists of a broad profile spanning from $600$ to $1000$ nm, with no highly dominant peaks, reflecting mainly the gain bandwidth of the Ti:sapphire crystal.
As the pump power increases and the system enters the QML regime, the spectral structure changes markedly, reflecting the onset of mode competition and frustration driven by nonlinear interactions. 
In this regime, a variety of spectral configurations may arise, including competing dominant peaks or single peaks exhibiting strong intensity fluctuations, depending on the specific experimental conditions \cite{ref28}. 
For the homemade Ti:sapphire laser, the QML regime is characterized by two dominant spectral peaks centered around wavelengths $780$ nm and $781$ nm (inset, Fig. \ref{fig:fig_2}(a)). 
In contrast, in the commercial Mira system, the QML regime is observed as a single spectral peak with strong intensity fluctuations, indicative of intermittent behavior, with a central wavelength around $798$ nm (inset, Fig. \ref{fig:fig_2}(b)).
These differences are due to variations in cavity design and operating conditions. 
Despite this, the qualitative features of the QML regime, namely, strong fluctuations and nonstationary spectral dynamics, are preserved across both systems and remain consistent with the intermediate phase identified in the optical phase diagram (Fig. \ref{fig:fig_1}).

At higher pump powers, the system shows transitions into the SML regime. 
In this regime, the spectral bandwidth broadens compared to the QML regime, reaching approximately $20$ nm for the homemade laser and about $10$ nm for the commercial system.
In addition, strong intensity fluctuations are suppressed as the system enters a more stable configuration. 
It is important to note that, although a complete time-domain characterization is important for fully map the laser dynamics, the RSB framework relies strictly on the statistical correlation of spectral intensity fluctuations. Within this framework, the clear correspondence between spectral features and dynamical regimes, as summarized in Fig. \ref{fig:fig_2} and mapped onto the phase diagram in Fig. \ref{fig:fig_1}, establishes the experimental basis for the statistical analysis used here.

\subsection*{Statistical tools}

RSB in photonic systems is investigated through the analysis of normalized correlations among intensity fluctuations in the optical spectra, quantified by the Parisi overlap parameter. RSB relies on comparing multiple realizations (replicas) of a system prepared under identical macroscopic conditions. In our experiment, each realization corresponds to a single optical spectrum recorded sequentially under stationary operating conditions. Because the spectrometer integration time is much longer than the cavity round-trip time, each measurement represents a time-integrated spectral realization rather than a time-resolved pulse, so the overlap parameter $q_{\alpha\beta}$ quantifies the similarity between successive time-integrated acquisitions.
For two distinct spectra $\alpha$ and $\beta$, the overlap parameter $q_{\alpha\beta}$ is defined as \cite{ref14, ref37}
%

\begin{equation}
    q_{\alpha\beta}=\frac{\sum_{_{k}}\Delta I_{\alpha}\left( \omega_{k} \right)\Delta I_{\beta}\left( \omega_{k} \right)}{\sqrt{\sum_{k}\left[ \Delta I_{\alpha}\left( \omega_{k} \right) \right]^{2}\sum_{k}\left[ \Delta I_{\beta}\left( \omega_{k} \right) \right]^{2}}},
    \label{eq:equation_1}
\end{equation}

\noindent where $\alpha, \beta=[1, 2,\ldots, N_R]$ are the spectral replicas labels, with $N_R={1\times10}^4$ the total number of replicas (spectral acquisitions) for both systems. 
The mean spectral intensity at a given frequency $\omega_k$ is defined as $\left\langle I\left( \omega_{k} \right) \right\rangle = \sum_{\alpha=1}^{N_{R}}I_{\alpha}\left( \omega_{k} \right)/N_{R}$, from which the intensity fluctuations are obtained as $\Delta I_{\alpha}\left( \omega_{k} \right)=I_{\alpha}\left( \omega_{k} \right)-\left\langle I\left( \omega_{k} \right)  \right\rangle$.
By evaluating $q_{\alpha\beta}$ over all pairs of distinct replicas ($N_R(N_R-1)/2$), the probability distribution $P(q)$ can be constructed. 
A distribution $P(q)$ exhibiting a single maximum at $q=0$ indicates replica-symmetric behavior, corresponding to paramagnetic-like or ferromagnetic-like phases, whereas the appearance of side maxima around $q\approx\pm1$  is associated with the RSB regime, characteristic of a glassy phase \cite{ref13}.

While the Parisi overlap captures correlations between replicas, spectral correlations of two distinct frequencies $\omega_{k_i}$ and $\omega_{k_j}$ within a single spectrum $\alpha$ can be further analyzed using the Pearson correlation coefficient \cite{ref30, ref38}

\begin{equation}
    C_{{k_i}{k_j}}=\frac{\sum_{_{\alpha}}\Delta I_{\alpha}\left( \omega_{k_i} \right)\Delta I_{\alpha}\left( \omega_{k_j} \right)}{\sqrt{\sum_{\alpha}\left[ \Delta I_{\alpha}\left( \omega_{k_i} \right) \right]^{2}\sum_{\alpha}\left[ \Delta I_{\alpha}\left( \omega_{k_j} \right) \right]^{2}}}.
    \label{eq:equation_2}
\end{equation}

Even though analogous, Parisi and Pearson parameters probe different aspects of the system.
$q_{\alpha\beta}$ measures correlations between different replicas by summing over frequencies, whereas the Pearson coefficient evaluates correlations between different frequency components within a single realization by summing over replicas. 
A value of $C_{k_ik_j}$ close to zero indicates that fluctuations at the two frequencies are essentially uncorrelated. 
In contrast, positive (negative) values of $C_{k_ik_j}$ reveal correlated (anti-correlated) behavior, meaning that intensity fluctuations at one frequency tend to occur with the same (opposite) sign at the other frequency.

\section*{Discussion}

\begin{figure}[ht!]
\centering
\includegraphics[width=\textwidth]{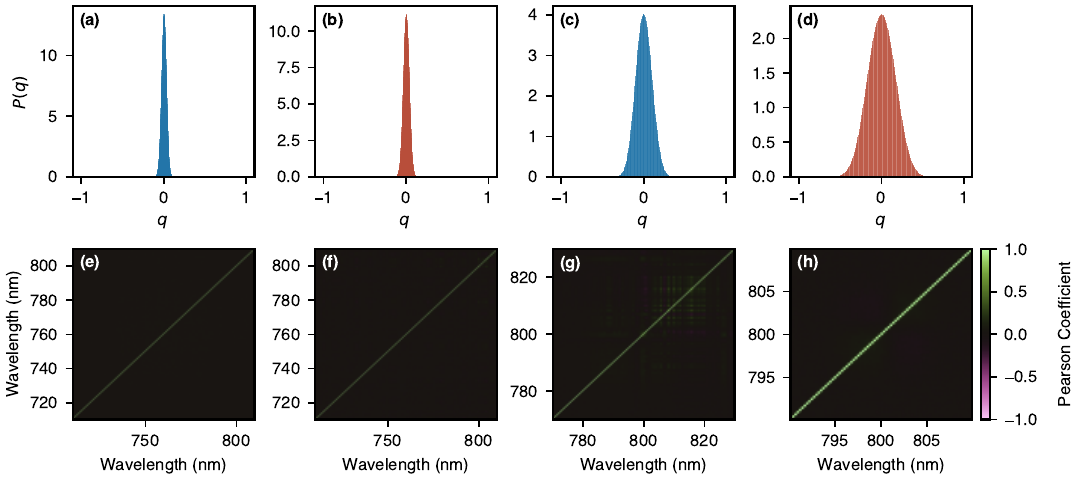}
\caption{Statistical characterization of the replica-symmetric regimes. \textbf{Top row} (a–d): Parisi overlap distributions $P(q)$ for the homemade Ti:sapphire laser (blue) and the commercial Coherent Mira (red). Panels (a,b) correspond to the ASE regime, while (c,d) correspond to the SML regime, for pump powers indicated in the text. In all cases, the distributions exhibit a single maximum centered at $q=0$, consistent with replica-symmetric behavior. $P(q)$ is normalized to unit area. \textbf{Bottom row} (e–h): Pearson correlation coefficients matrices for the corresponding regimes. In the ASE regime, the Pearson matrix shows uncorrelated spectral fluctuations, whereas the SML regime presents weak correlations associated with the formation of phase-locked modes.}
\label{fig:fig_3}
\end{figure}

The overlap distributions $P(q)$ for the ASE and SML regimes in both Ti:sapphire systems are shown in the top row of Fig. \ref{fig:fig_3}, where blue corresponds to the homemade laser and red to the commercial Mira. 
Panels (a,b) correspond to the ASE regime, while panels (c,d) correspond to the SML regime.
The pump powers are the same as those for the spectra shown in Fig. \ref{fig:fig_2}. In all cases, $P(q)$ exhibits a single maximum at $q=0$, indicating replica-symmetric behavior.

Although both ASE and SML regimes are replica-symmetric, their physical origins are fundamentally different. 
In the ASE regime, light arises from spontaneous emission within the gain medium, which is amplified by the medium itself. 
This leads to a rapidly changing optical field with a broad spectrum and low temporal coherence. 
Since the process arises from independent random emission events, the intensity fluctuations behave as uncorrelated noise with approximately Gaussian statistics. 
This reflects the absence of collective mode interactions, which is consistent with replica-symmetric, paramagnetic-like behavior.

In contrast, the SML corresponds to a stable operation sustained by a balance between gain, dispersion, and nonlinear effects.
Nonlinear interactions distribute the intracavity energy over a broad set of longitudinal modes around the central emission of the gain medium, synchronizing their phases into a well-ordered state. 
Analogously to a ferromagnet where spins are aligned, these modes synchronize their phases, minimizing losses and concentrating the total energy into a single, high-intensity temporal wavepacket. The system remains replica-symmetric because the mode dynamics converge to a unique global equilibrium configuration without frustration.

This distinction may become evident in the Pearson correlation maps shown in Fig. \ref{fig:fig_3}(e–h). 
In the ASE regime, the maps display negligible off-diagonal correlations, indicating that fluctuations at different frequencies are largely independent.
On the other hand, in the SML regime, weak but structured correlations may emerge (see Fig.\ref{fig:fig_3}(g,h)), reflecting the collective organization of modes associated with the mode-locking process. 
These results highlight that, although both regimes share the same replica-symmetric signature in P(q), their internal spectral dynamics are markedly different. 
A similar behavior has also been reported in ytterbium/erbium-based mode-locked fiber lasers \cite{ref28, ref29}, suggesting that this phenomenon may be more general.

\begin{figure}[ht!]
\centering
\includegraphics[width=120 mm]{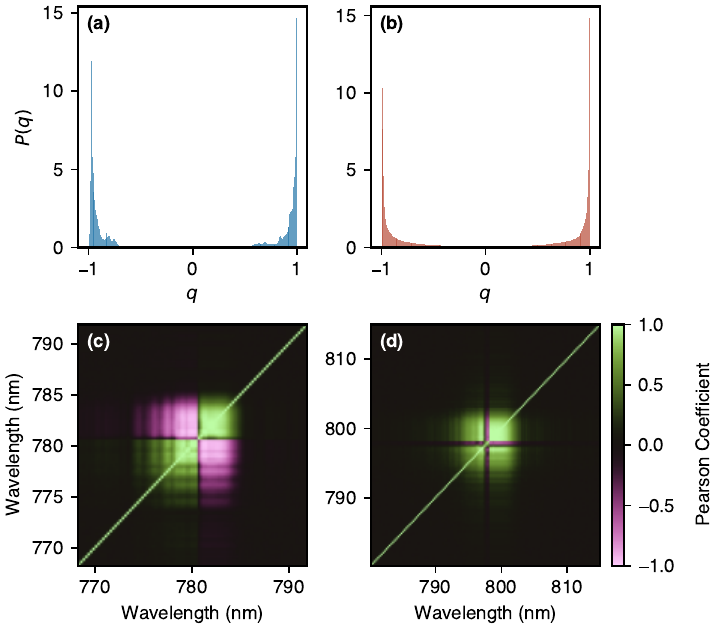}
\caption{Statistical signatures of the QML regime. Top row (a,b): Parisi overlap distributions $P(q)$ for the homemade Ti:sapphire laser (blue) and the commercial Coherent Mira system (red), respectively. In both cases, the distributions exhibit a clear bimodal structure with maxima at $q\approx\pm1$, indicating the presence of RSB. Bottom row (c,d): Corresponding Pearson correlation coefficient maps $C_{k_ik_j}$. The maps reveal the coexistence of correlated and anticorrelated spectral regions, reflecting the competition between spectral modes characteristic of the QML regime. The pump powers corresponding to each case are indicated in the main text.}
\label{fig:fig_4}
\end{figure}

A markedly different statistical behavior emerges in the intermediate QML regime, as shown in Fig. \ref{fig:fig_4}. 
Panels (a,b) correspond to the overlap distribution $P(q)$ for the homemade Ti:sapphire laser (blue) and the commercial Mira system (red), for the same pumps described in Fig. \ref{fig:fig_2}. In contrast to the ASE and SML regimes, $P(q)$ exhibits a clear bimodal structure, with maxima located at $q\approx\pm1$  ($\left|q_{max}\right|\ \approx1$). This behavior is the hallmark of RSB and reflects the existence of frustrated interactions. In the QML regime, frustration emerges from competition among spectral modes for the finite gain available in the medium.
%

The Pearson correlation maps shown in Fig. \ref{fig:fig_4}(c,d) provide further insight into the behavior of this phase. 
The QML regime exhibits strong correlations and anticorrelations among spectra. 
The correlation pattern evolves with pump power, reflecting the system’s access to different dynamical configurations within the QML phase. This behavior indicates the coexistence of competing spectral modes, with energy being dynamically redistributed among different frequency components.

To show that the systems go through three different regimes of operation, we plot the absolute value of $q_{\max}$ as a function of pump power. Fig. 5(a) shows the curve for the homemade Ti:sapphire laser, while Fig. 5(b) shows the corresponding results for the commercial system.

The transition points between dynamical regimes depend on the cavity parameters. For the homemade Ti:sapphire laser, the onset of the QML regime occurs at a pump power of approximately $0.7$ W, whereas in the commercial system, it occurs around $2.9$ W. In both cases, the transition from QML to SML takes place at pump powers of about $4.8$ W. These thresholds are consistent with the spectral evolution across the ASE, QML, and SML regimes, providing a way to identify the different dynamical states. 

For the homemade laser, $q_{\max}$ remains stable throughout the QML, whereas for the commercial one, $q_{\max}$ decreases in some parts of this region, returning toward near zero before the SML transition. We attribute this difference to the presence of additional frequency-selective elements in the commercial cavity (see cavity design in the Experimental Details section). Specifically, the cavity mirrors are designed to restrict the supported wavelength range, and a birefringent filter (BRF, or Lyot filter) further selects a narrower portion within this range \cite{ref36}. These elements reduce the spectral bandwidth of participating modes, which we suggest alters the structure of the four-wave mixing interaction network governing modal correlations, resulting in a decrease of  $q_{\max}$. 
Nevertheless, non-Gaussian fluctuations persist throughout the QML region of the commercial system, indicating that the local nonlinear dynamics of the central modes remain intact. These results suggest that an interplay can be present between frequency-selective components and modal coupling, calling for further investigation.

\begin{figure}[ht!]
\centering
\includegraphics[width=\textwidth]{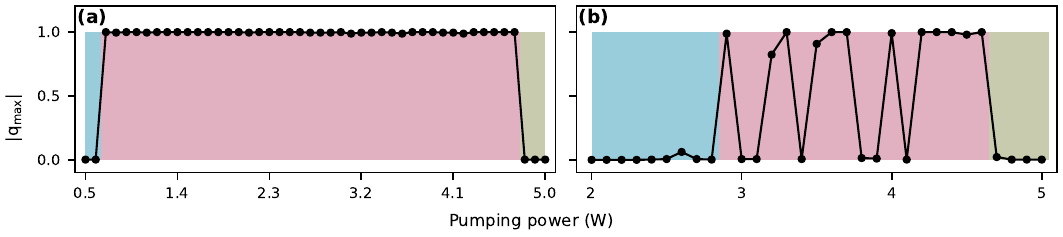}
\caption{Absolute maximum of the Parisi overlap parameter, $q_{\max}$, as a function of pump power for (a) the homemade and (b) the commercial Ti:sapphire laser. The blue, red, and green shaded regions indicate the ASE, QML, and SML regimes, as in the phase diagram of Fig. \ref{fig:fig_1}. The onset of the QML regime occurs at approximately $0.6$ W and $2.9$ W for the homemade and commercial cavities, respectively, while the transition to the SML regime takes place at approximately $4.8$ W in both cases.}
\label{fig:fig_5}
\end{figure}

Despite this, a consistent picture of the QML regime can be provided, in which strong fluctuations, mode competition, and frustration collectively give rise to replica symmetry breaking. The combined analysis of the Parisi overlap distributions and Pearson correlation maps reveal how nonlinear interactions govern the statistical dynamics of this regime, clearly distinguishing it from the ASE and SML. Moreover, the results are consistent with previous observations in ytterbium-based mode-locked fiber lasers \cite{ref28} and indicate that the same statistical signatures are also present in Ti:sapphire systems.

\section*{Experimental Details}

\begin{figure}[ht!]
\centering
\includegraphics[width=\textwidth]{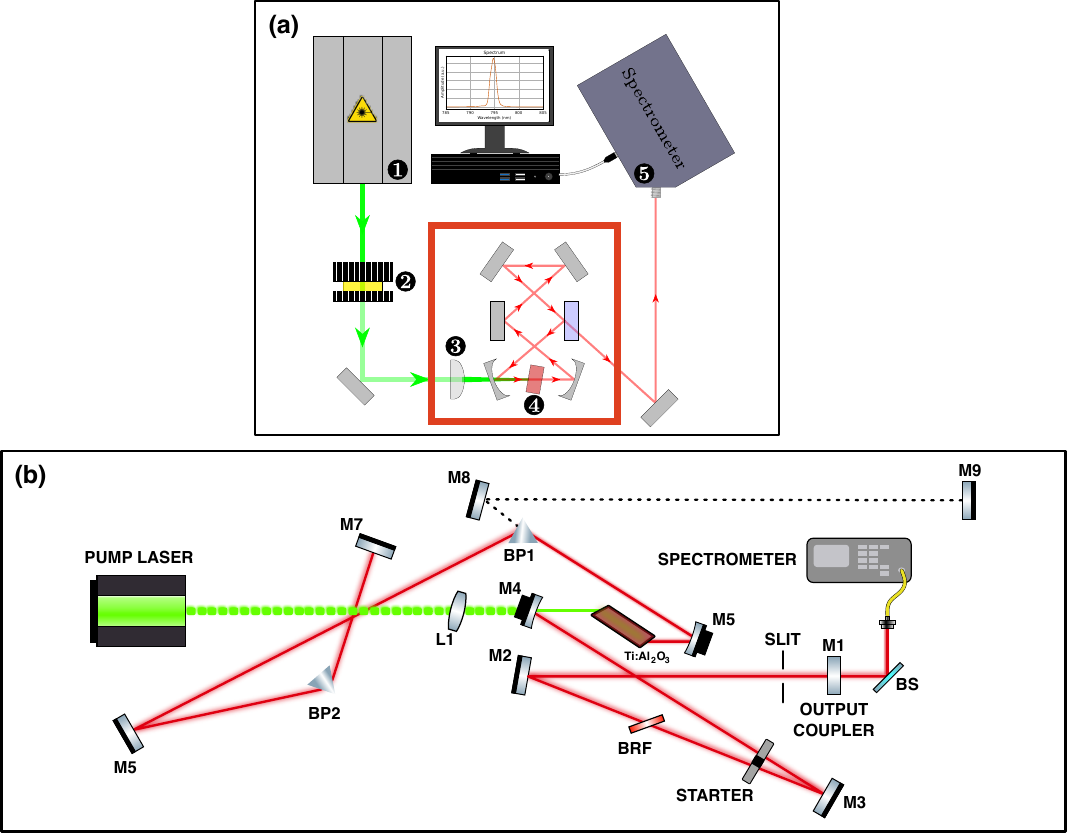}
\caption{Schematic of the Ti:sapphire laser cavities used in this work: (a) Homemade cavity, which main components are: (1) pump laser, (2) half-wave plate, (3) focusing lens, (4) laser cavity, comprising the gain crystal and cavity mirrors, and (5) spectrometer. (b) Commercial Mira Optima 900-F cavity (adapted from user's manual \cite{ref36}). M1–M9 are mirrors, L1 is a focusing lens, BP1 and BP2 are Brewster prisms, and the gain medium is a $\text{Ti:Al}_2\text{O}_3$ crystal. BS is a beamsplitter that directs the output light to a spectrometer. BRF is a birefringent (Lyot) filter. The slit introduces higher round-trip losses for larger beam profiles, which favors mode-locked over low-excitation operation. The starter initiates mode-locked operation via rapid, small changes in the cavity length. M8 and M9 constitute an auxiliary cavity used during initial alignment (dashed lines), formed by removing BP1 from the beam path.}
\label{fig:fig_6}
\end{figure}

Fig. 6 shows a schematic of the Ti:sapphire laser cavities used in this work, where Fig. 6(a) corresponds to the homemade cavity and Fig. 6(b) to the commercial one. For both cavities, it is possible to access the ASE, QML, and SML regimes by controlling the pump power.

The homemade cavity consists of a 2.9-mm-thick Ti:sapphire crystal acting as the gain medium, pumped by a continuous-wave laser at 532 nm (Spectra-Physics Millennia eV 55), with pump power controlled in the range from 0.50 W to 5.00 W. A half-wave plate is used to adjust the polarization of the pump beam, which is subsequently focused onto the crystal by a lens with a 30 mm focal length. The cavity is formed by two plano-concave mirrors (R = 30 mm) and dispersive dielectric plane mirrors. The output beam is extracted through a $5\%$ transmission output coupler, with a total cavity length of approximately 36 cm, corresponding to a repetition rate of about 840 MHz. The mode-locking is achieved through the Kerr-lens effect in the Ti:sapphire crystal, without the use of an external saturable absorber. The optical Kerr effect induces an intensity-dependent refractive index ($\Delta n \propto I$), causing the high-intensity portion of the intracavity beam to experience stronger self-focusing than the low-intensity background. By adjusting the cavity close to the stability limit, this nonlinear self-focusing produces lower round-trip losses for short pulses than for continuous-wave operation, mimicking the effect of saturable absorption and initiating and sustaining Kerr-lens mode-locking. In the homemade cavity, this is obtained by optimizing the spatial overlap between the focused pump beam and the intracavity mode inside the Ti:sapphire crystal, followed by fine adjustment of the cavity alignment. Intracavity dispersion is compensated using dispersive dielectric mirrors to support stable ultrashort pulse generation.

The emitted radiation is directly coupled to a spectrometer (Ocean Optics HR4000CG-UV-NIR), which has a nominal operating range of $200-1100$ nm, with an integration time of $6$ ms, and an optical resolution of $\approx 0.25$ nm. 
Consequently, each acquired spectrum corresponds to the emission integrated over approximately multiple cavity round trips in the pulsed regime, and therefore constitutes a time-integrated spectral realization used in the statistical analysis. For the statistical analysis, the spectral window was restricted to exclude wavelengths dominated by the detector noise background, as these contribute purely with uncorrelated fluctuations to the overlap statistics. This results in an analysis performed exclusively within the spectral range shown in Fig.~\ref{fig:fig_2}. The spectrometer is connected to a computer, which performs the data acquisition reported here in this work. For each pump power, a sequence of $10^4$ optical spectra is acquired using a Python-based script that directly interfaces with the spectrometer. This procedure allows for fast recorded measurements, with negligible dead time between acquisitions when compared to spectrometer integration time.

Regarding the commercial Ti:sapphire laser system, it relies on a similar physical operating principle but incorporates a more integrated optical design, with reduced flexibility for direct control of intracavity parameters. Fig. 6(b) is adapted from the laser manual, where more details can be found \cite{ref36}. The $\text{Ti:Al}_2\text{O}_3$ crystal is pumped by a Coherent Verdi laser at 532 nm. The output beam, leaving through the output coupler, is directed to the same spectrometer used for the homemade cavity. The Mira Optima 900-F also operates by Kerr-lens mode locking. Intracavity group-velocity dispersion is compensated by a Brewster prism pair (BP1-BP2), with the insertion of the second prism adjusted to optimize the cavity group-velocity dispersion and the output pulse duration. The birefringent filter (BRF) provides continuous wavelength tuning, while the slit introduces wavelength-dependent losses that help define the operating bandwidth. Mode locking is initiated by the built-in starter, which introduces a rapid perturbation of the cavity length through a galvanometric actuator, allowing the laser to transition from continuous-wave to stable Kerr-lens mode-locked operation. The commercial cavity is constructed in a way that favors single-mode operation, suppressing the dynamics of competing modes and playing an important role in the dynamical behavior observed in the QML regime.

\section*{Conclusion}

In summary, this work provides clear experimental evidence of the RSB phenomena in Ti:sapphire lasers operating in the QML regime. By analyzing the statistical correlations of spectral intensity fluctuations across both a homemade cavity and a commercial system, three distinct optical phases were identified using Parisi overlap distributions $P(q)$ and Pearson correlation matrices. While the low-excitation ASE and SML regimes display replica-symmetric behavior with single-peaked distributions centered at $q = 0$, the intermediate QML regime exhibits a bimodal $P(q)$ distribution with peaks at $q \approx \pm 1$, confirming the presence of an RSB phase. These statistical signatures reflect fundamental differences in underlying modal interactions: ASE represents uncorrelated noise analogous to a paramagnetic state, whereas SML reflects a globally synchronized state akin to a ferromagnet. In contrast, the QML regime is driven by frustrated mode competition for gain, resulting in a photonic glassy-like phase characterized by coexisting correlated and anticorrelated spectral fluctuations. Demonstrating this glassy behavior in Ti:sapphire systems confirms that RSB in the QML regime is not confined to fiber lasers, supporting it as a general property of multimode laser systems prior to stable pulse formation.

\section*{Data Availability}
The data that support the findings of this study are available from the corresponding author upon reasonable request.

\bibliography{references}

\section*{Funding}

We thank the financial support from the agencies Coordenação de Aperfeiçoamento de Pessoal de Nível Superior (CAPES) Conselho Nacional de
Desenvolvimento Científico e Tecnológico (CNPq), National Quantum Information Institute (INCT-IQ, No. 410 465469/2014-0)/CNPq, Army Research Office (Grant No. W911NF-23-1-0287), and the National Photonics Institute (INCT-INFo, No. 409174/2024-6)/CNPq, FAPESP (2021/06535-0).

\section*{Author contributions statement}

Experimental data acquisition: R.V. M. A. F., I. C. N. and G. S. B.; Experimental supervision: C. B. A., A. C. A. S. and M. G. H. M.; Data analysis and figures: N. P. A., R. V. M. A. F. and I. C. N.; Overall coordination: A. C. A. S.. All authors reviewed the manuscript.

\section*{Additional information}

The authors declare no competing interests.

\end{document}